\documentclass[thmsa,onecolumn,11pt]{article}
\usepackage{sw20lart}

\input{tcilatex}
\begin{document}

\vspace*{0in}

\begin{center}
{\baselineskip 25pt }
\end{center}

{\Large \textbf{Branching Ratio and CP-asymmetry for}}{\Huge \ }$%
B\rightarrow 1^{1}P_{1}${\Huge \ }$\gamma ${\Large \textbf{\ decays}}

{\Large \ }

\begin{center}
\vspace{1.2cm} \centerline{\bf M. Jamil Aslam}

\vspace{0.5cm} {\small \textit{National Centre for Physics and Department of
Physics, Quaid-i-Azam University, Islamabad \footnote{\textit{E-mail:
jamil@ncp.edu.pk}}} \vspace{0.5cm} }

{\small \centerline{\bf Riazuddin} \vspace{0.5cm} \textit{National Centre
for Physics, Quaid-i-Azam University Campus, Islamabad \footnote{\textit{%
E-mail: riazuddin@ncp.edu.pk}}} }

{\small \vspace{0.5cm} }
\end{center}


\bigskip 

We calculate the branching ratios for $B_{d}^{0}\rightarrow \left(
b_{1},h_{1}\right) \gamma $ at next-to-leading order (NLO) of $\alpha _{s}$
where $b_{1}$ and $h_{1}$ are the corresponding radially excited axial
vector mesons of $\rho $ and $\omega $ respectively. Using the $SU\left(
3\right) $ symmetry for the form factor, the branching ratio for $%
B_{d}^{0}\rightarrow \left( b_{1},h_{1}\right) \gamma $ is expressed in
terms of the branching ratio of the $B_{d}^{0}\rightarrow K_{1}\gamma $ and
it is found to be $\mathcal{B}\left( B_{d}^{0}\rightarrow b_{1}\gamma
\right) =0.71\times 10^{-6}$ and $\mathcal{B}\left( B_{d}^{0}\rightarrow
h_{1}\gamma \right) =0.74\times 10^{-6}$. We also calculate direct CP
asymmetry for these decays and find, in confirmity with the observations
made in the literature, that the hard spectator contributions significantely
reduces the asymmetry arising from the vertex corrections alone. The value
of CP-asymmetry is 10\% and is negative like $\rho $ and $\omega $ in the
Standard Model.

\newpage


\paragraph{1. Introduction.}

%
The Flavor-Changing-Neutral-Current (FCNC) processes which cause $%
b\rightarrow s\gamma $ and $b\rightarrow d\gamma $ decays may contain new
physics (NP) effects through penguin amplitudes. As the SM effects represent
the back ground when we search for NP effects, we shall compute these
effects. In doing so, we can understand the sensitivity of each NP search.

The first experimental evidence of this FCNC transition process in $B$ decay
was observed about a decade ago, where the inclusive process $b\rightarrow
s\gamma $ and exclusive process $B\to K^{*}\gamma $ were detected, and their
branching ratios were measured\cite{Ammar:1993sh,Alam:1994aw}. On the other
hand, the expected branching ratio for $b\rightarrow d\gamma $ is suppressed
by $\mathcal{O}\left( 10^{-2}\right) $ with respect to the $b\rightarrow
s\gamma $, because of the Cabbibo-Kobayashi-Masukawa quark mixing matrix
factor (CKM). The world average for $b\rightarrow d$ penguin decays are
given as follow\cite{HFAG} 
\begin{eqnarray}
\mathcal{B}\left( B^{0}\to \rho ^{0}\,\gamma \right) &=&\left( 0.38\pm
0.18\right) \times 10^{-6}  \label{Branchingratio1} \\
\mathcal{B}\left( B^{0}\to \omega \,\gamma \right) &=&\left(
0.54_{-0.21}^{+0.23}\right) \times 10^{-6}  \nonumber \\
\mathcal{B}\left( B^{+}\to \rho ^{+}\,\gamma \right) &=&\left(
0.68_{-0.31}^{+0.36}\right) \times 10^{-6}  \nonumber
\end{eqnarray}

Theoretically, $B\rightarrow \left( \rho \text{,}\omega \right) \gamma $ are
widely studied both within and beyond the SM\cite{cluster1,cluster2}. Now
after the first measurment of BELLE for the decay $B\to K_{1}\gamma $, where 
$K_{1}$ are the higher resonaces of kaon\cite{Belle}, these higher states
become a subject of topical interest for the theoreticians. These decays
have been studied widely in the literature\cite{Safir,Lee,Leenew,Jamil}.
Recently, the leading twist LCDAs as well as the first few Gegenbauer
moments of $1^{1}P_{1}$ mesons, $b_{1}\left( 1235\right) $ and $h_{1}\left(
1170\right) $, which are the axial vector states of the $\rho $ and $\omega $
mesons have been studied\cite{Yang}. It is pointed out that these LCDAs are
not only important to explore the tensor-type new-physics in B decays but
also for $B\to 1^{1}P_{1}\gamma $ studies.

In this paper the branching ratio for $B_{d}^{0}\rightarrow \left( b_{1}%
\text{,}h_{1}\right) \gamma $ at NLO of $\alpha _{s}$ are calculated using
the LEET\ approach\cite{Dugan,Charles}. We follow the same frame work as
done by Ali et al.\cite{Alinew} for $B\rightarrow \left( \rho \text{,}\omega
\right) \gamma $, because $B_{d}^{0}\rightarrow \left( b_{1}\text{,}%
h_{1}\right) \gamma $ shares many things with it. The only difference is the
DA for the daughter meson. As $\left( b_{1}\text{,}h_{1}\right) $ is an
axial vector and is distinguished by vector by the $\gamma _{5}$ in the
gamma structure of DA and some non perturbative parameters. But the presence
of $\gamma _{5}$ does not alter the calculation, give the same result for
the perturbative part. The higher twist terms are also included through the
Gegenbauer moments in the Gegenbauer expansion.

At next-to-leading order of $\alpha _{s}$, $B\rightarrow \left( \rho \text{,}%
\omega \right) \gamma $ and $B_{d}^{0}\rightarrow \left( b_{1}\text{,}%
h_{1}\right) \gamma $ are characterized by the weak form factor and decay
constant, plugged by the common perturbative and kinematical factors. With $%
\mathcal{B}\left( B\rightarrow \left( \rho \text{,}\omega \right) \gamma
\right) $ at hand, we can say that the future experiment will check the
structure for $B_{d}^{0}\rightarrow \left( b_{1}\text{,}h_{1}\right) \gamma $%
.

\paragraph{2. Effective Hamiltonian.}

%
The effective Hamiltoinan for the radiative $b\to d\gamma $ decays
(equivalently $B_{d}^{0}\to b_{1}\gamma $ and $B_{d}^{0}\to h_{1}\gamma $
decays) is obtained from the Standard Model (SM) by integrating out the
heavy degrees of freedom (the top quark and $W^{\pm }$-bosons). The
resulting expression at the scale $\mu =O(m_{b})$, where~$m_{b}$ is the $b$%
-quark mass, is given by 
\begin{eqnarray}
\mathcal{H}_{\mathrm{eff}}^{b\to d} &=&\frac{G_{F}}{\sqrt{2}}\,\left\{
V_{ub}V_{ud}^{*}\,\left[ C_{1}^{(u)}(\mu )\,\mathcal{O}_{1}^{(u)}(\mu
)+C_{2}^{(u)}(\mu )\,\mathcal{O}_{2}^{(u)}(\mu )\right] \right.
\label{eq:eff-ham} \\
&&\qquad +\,V_{cb}V_{cd}^{*}\,\left[ C_{1}^{(c)}(\mu )\,\mathcal{O}%
_{1}^{(c)}(\mu )+C_{2}^{(c)}(\mu )\,\mathcal{O}_{2}^{(c)}(\mu )\right] 
\nonumber \\
&&\qquad -\,\left. V_{tb}V_{td}^{*}\,\left[ C_{7}^{\mathrm{eff}}(\mu )\,%
\mathcal{O}_{7}(\mu )+C_{8}^{\mathrm{eff}}(\mu )\,\mathcal{O}_{8}(\mu
)\right] +\ldots \right\} ,  \nonumber
\end{eqnarray}
where~$G_{F}$ is the Fermi coupling constant and only the dominant terms are
shown. The operators~$\mathcal{O}_{1}^{(q)}$ and~$\mathcal{O}_{2}^{(q)}$, $%
(q=u,c)$, are the standard four-fermion operators and~$\mathcal{O}_{7}$ and~$%
\mathcal{O}_{8}$ are the electromagnetic and chromomagnetic penguin
operators, respectively. Their explict expressions are 
\begin{equation}
\mathcal{O}_{1}^{(q)}=(\bar{d}_{\alpha }\gamma _{\mu }(1-\gamma
_{5})q_{\beta })\,(\bar{q}_{\beta }\gamma ^{\mu }(1-\gamma _{5})b_{\alpha
}),\qquad \mathcal{O}_{2}^{(q)}=(\bar{d}_{\alpha }\gamma _{\mu }(1-\gamma
_{5})q_{\alpha })\,(\bar{q}_{\beta }\gamma ^{\mu }(1-\gamma _{5})b_{\beta }),
\label{eq:four-Fermi}
\end{equation}
\begin{equation}
\mathcal{O}_{7}=\frac{em_{b}}{8\pi ^{2}}\,(\bar{d}_{\alpha }\sigma ^{\mu \nu
}(1+\gamma _{5})b_{\alpha })\,F_{\mu \nu },\qquad \mathcal{O}_{8}=\frac{%
g_{s}m_{b}}{8\pi ^{2}}\,(\bar{d}_{\alpha }\sigma ^{\mu \nu }(1+\gamma
_{5})T_{\alpha \beta }^{a}b_{\beta })\,G_{\mu \nu }^{a}.
\label{eq:mag-penguin}
\end{equation}
Here,~$e$ and~$g_{s}$ are the electric and colour charges, $F_{\mu \nu }$
and~$G_{\mu \nu }^{a}$ are the electromagnetic and gluonic field strength
tensors, respectively, $T_{\alpha \beta }^{a}$ are the colour $SU(N_{c})$
group generators, and the quark colour indices~$\alpha $ and~$\beta $ and
gluonic colour index~$a$ are written explicitly. Note that in the operators~$%
\mathcal{O}_{7}$ and~$\mathcal{O}_{8}$ the $d$-quark mass contributions are
negligible and therefore omitted. The coefficients~$C_{1}^{(q)}(\mu )$ and~$%
C_{2}^{(q)}(\mu )$ in Eq.~(\ref{eq:eff-ham}) are the usual Wilson
coefficients corresponding to the operators~$\mathcal{O}_{1}^{(q)}$ and~$%
\mathcal{O}_{2}^{(q)}$, while the coefficients~$C_{7}^{\mathrm{eff}}(\mu )$
and~$C_{8}^{\mathrm{eff}}(\mu )$ include also the effects of the QCD penguin
four-fermion operators which are assumed to be present in the effective
Hamiltonian~(\ref{eq:eff-ham}) and denoted by ellipses there. For details
and numerical values of these coefficients, see~Ref.\cite{Buchalla:1996vs}
and also references therein. We use the standard Bjorken-Drell convention~%
\cite{Bjorken:1965} for the metric and the Dirac matrices; in particular $%
\gamma _{5}=i\gamma ^{0}\gamma ^{1}\gamma ^{2}\gamma ^{3}$, and the totally
antisymmetric Levi-Civita tensor $\varepsilon _{\mu \nu \rho \sigma }$ is
defined as $\varepsilon _{0123}=+1$. A point to note is that the three CKM
factors shown in~$\mathcal{H}_{\mathrm{eff}}^{b\to d}$ are of the same order
of magnitude and, hence, the matrix elements in the decays $b\to d\gamma $
and $B_{d}^{0}\to (b_{1},h_{1})\gamma $ have non-trivial dependence on the
CKM parameters. This is not the case of $b\to s\gamma $ decay (equivalently
the $B\to K_{1}\gamma $ decays), the effective Hamiltonian~$\mathcal{H}_{%
\mathrm{eff}}^{b\to s}$ describing the $b\to s$ transition can be obtained
by the replacement of the quark field~$d_{\alpha }$ by~$s_{\alpha }$ in all
the operators in Eqs.~(\ref{eq:four-Fermi}) and~(\ref{eq:mag-penguin}) and
by replacing the CKM factors $V_{qb}V_{qd}^{*}\to V_{qb}V_{qs}^{*}$ ($%
q=u,c,t $) in $\mathcal{H}_{\mathrm{eff}}^{b\to d}$~(\ref{eq:eff-ham}).
Noting that among the three factors~$V_{qb}V_{qs}^{*}$, the combination~$%
V_{ub}V_{us}^{*} $ is CKM suppressed, the corresponding contributions to the
decay amplitude can be safely neglected.

\paragraph{3. Theoretical framework for the $B\to 1^{1}P_{1}\gamma $ decays.}

\label{sec:Theory} The matrix element for the $B_{d}^{0}\to 1^{1}P_{1}\gamma 
$ ($1^{1}P_{1}=b_{1},h_{1}$) decays, we need to calculate the matrix
elements $\langle 1^{1}P_{1}\gamma |\mathcal{O}_{i}|B\rangle $, where~$%
\mathcal{O}_{i}$ are the operators appearing in~$\mathcal{H}_{\mathrm{eff}%
}^{b\to s}$ and~$\mathcal{H}_{\mathrm{eff}}^{b\to d}$. At the leading order
in~$\alpha _{s}$, this involves only the operators~$\mathcal{O}_{7}$,~$%
\mathcal{O}_{1}^{(u)}$ and~$\mathcal{O}_{2}^{(u)}$. The contribution from $%
\mathcal{O}_{7}$ is termed as the long-distance contribution characterized
by the top quark induced amplitude, where $\mathcal{O}_{1}^{(u)}$ and~$%
\mathcal{O}_{2}^{(u)}$ corresponds to the short distance contributions and
it includes the penguin amplitude for the $u$ and $c\,$quark intermediate
states and also the so-called weak annihilation and $W$-exchange
contributions. There is also some contribution from annihilation penguin
diagrams, which, however, are small. For detailed discussion about these
kind of topoligies for $B\to V\gamma $ decays and references to earlier
papers, see Ref.~\cite{Grinstein:2000pc}. Recently it has been shown that
for the higher kaon resonances $K_{1}$, the branching ratio for $%
B\rightarrow K_{1}\gamma $ has negligable dependence on such kind of
annihilation topologies\cite{jamilnew}.

To calculate $O(\alpha _{s})$ corrections, all the operators listed in~(\ref
{eq:four-Fermi}) and~(\ref{eq:mag-penguin}) have to be included. QCD
factorization \cite{Beneke:1999br} is most convenient framework to carry out
these calcuations. This allows to express the hadronic matrix elements in
the schematic form: 
\begin{equation}
\langle 1^{1}P_{1}\gamma |\mathcal{O}_{i}|B\rangle =F^{B\to 1^{1}P_{1}}%
\mathcal{T}_{i}^{I}+\int \frac{dk_{+}}{2\pi }\int\limits_{0}^{1}du\,\phi
_{B,+}(k_{+})T_{i}^{II}(k_{+},u)\phi _{\perp }^{\left( 1^{1}P_{1}\right)
}(u),  \label{eq:bbnsfact}
\end{equation}
where $F^{B\to 1^{1}P_{1}}$ are the transition form factors defined through
the matrix elements of the operator~$\mathcal{O}_{7}$. $\phi _{B,+}(k_{+})$
is the leading-twist $B$-meson wave-function with $k_{+}$ being a light-cone
component of the spectator quark momentum, $\phi _{\perp }^{\left(
1^{1}P_{1}\right) }(u)$ is the leading-twist light-cone distribution
amplitude (LCDA) of the transversely-polarized axial-vector meson, and~$u$
is the fractional momentum of the vector meson carried by one of the two
partons. The expressions for these wavefunctions are given in Ref. \cite
{Lee, Jamil}, where it was poined out that vector and axial vector mesons
are distinguished by $\gamma _{5}$ in the gamma structure of the decay
amplitude and some non perturbative parameters. The quantities~$\mathcal{T}%
_{i}^{I}$ and~$T_{i}^{II}$ are the hard-perturbative kernels calculated to
order~$\alpha _{s}$, with the latter containing the so-called hard-spectator
contributions. The factorization formula~(\ref{eq:bbnsfact}) holds in the
heavy quark limit, i.e., to order~$\Lambda _{\mathrm{QCD}}/M_{B}$. This
factorization framework has been used to calculate the branching fractions
and related quantities for the decays $B\to K^{*}\gamma $~\cite
{Ali:2001ez,Beneke:2001at,Bosch:2001gv} and $B\to \rho \gamma $~\cite
{Ali:2001ez,Bosch:2001gv} and for $B\to K_{1}\gamma $~\cite{Lee,Leenew,Jamil}%
. The isospin violation in the $B\to K^{*}\gamma $ decays in this framework
have also been studied~\cite{Kagan:2001zk}. Very recently, the
hard-spectator contribution arising from the chromomagnetic operator~$%
\mathcal{O}_{8}$ have also been calculated in next-to-next-to-leading order
(NNLO) in $\alpha _{s}$ showing that the spectator interactions factorize in
the heavy quark limit~\cite{Descotes-Genon:2004hd}. However, the numerical
effect of the resummed NNLO contributions is marginal and we shall not
include this in our update.

It is shown that the the extra $\gamma _{5}$ in the DA of axial vector meson
in comparsion to the vector meson does not alter the calculation, giving the
same result for the perturbative part. As for the non-perturbative
parameters, the decay constant is most important. The LCDA for $b_{1}$ and $%
h_{1}$ meson has recently been calculated in \cite{Yang}. The transverse
decay constant of these mesons as well as the first few Gagenbaur moments of
leading twist LCDA are calculated by using QCD sum rule techinque. Their
numerical values are given in Table 1.

In what follows we shall use the notations and results from Ref.~\cite{Jamil}%
, to which we refer for detailed derivations for $B\to K_{1}\gamma $ decay.
The final state $K_{1}$ is also the axial vector meson like $b_{1}$ and $%
h_{1}$ mesons. The only difference is in the quark content and we have to
change the $s$ quark with $d$ quark every where in the calculation. The
branching ratio of the $B_{d}^{0}\to \left( b_{1}\text{, }h_{1}\right) $%
decay corrected to $O(\alpha _{s})$ can be written as follows~\cite{Jamil}:

\begin{eqnarray}
\mathcal{B}_{\mathrm{th}}(B_{d}^{0}\to \left( b_{1}\text{, }h_{1}\right)
\gamma ) &=&\tau _{B}\,\Gamma _{\mathrm{th}}(B_{d}^{0}\to \left( b_{1}\text{%
, }h_{1}\right)  \nonumber \\
&=&\tau _{B}\,\frac{G_{F}^{2}\alpha |V_{tb}V_{td}^{*}|^{2}}{32\pi ^{4}}%
\,m_{b,\mathrm{pole}}^{2}\,M^{3}\,\left[ \xi _{\perp }^{(b_{1}\text{, }%
h_{1})}\right] ^{2}\left( 1-\frac{m_{(b_{1}\text{, }h_{1})}^{2}}{M^{2}}%
\right) ^{3}\left| C_{7}^{(0)\mathrm{eff}}+A^{(1)}(\mu )\right| ^{2} 
\nonumber \\
&&  \label{branching1}
\end{eqnarray}
where~$G_{F}$ is the Fermi coupling constant, $\alpha =\alpha (0)=1/137$ is
the fine-structure constant, $m_{b,\mathrm{pole}}$ is the pole $b$-quark
mass, $M$~and $m_{(b_{1}\text{, }h_{1})}$ are the $B$- and axial
vector-meson masses, and~$\tau _{B}$ is the lifetime of the~$B^{0}$- or $%
B^{+}$-meson. The value of these constants is used from\cite{Alinew,Jamil}
and are collected in Table 1, for the numerical analysis. For this study, we
consider $\xi _{\perp }^{(b_{1}\text{, }h_{1})}$ as a free parameter and we
will extract its value from the current experimental data on $B\to
K_{1}\gamma $ decays because $K_{1}$ is also an axial vector meson. This is
in anology with the calculation done for the branching ratio of $B\to \left(
\rho \text{, }\omega \right) \gamma $ in terms of the branching ratio of $%
B\to K^{*}\gamma $ by Ali et al.\cite{Alinew}.

The function~$A^{(1)}$ in Eq.~(\ref{branching1}) can be decomposed into the
following three components: 
\begin{equation}
A^{(1)}(\mu )=A_{C_{7}}^{(1)}(\mu )+A_{\mathrm{ver}}^{(1)}(\mu )+A_{\mathrm{%
sp}}^{(1)K_{1}}(\mu _{\mathrm{sp}})~.
\end{equation}
Here, $A_{C_{7}}^{(1)}$ and $A_{\mathrm{ver}}^{(1)}$ are the $O(\alpha _{s})$
(i.e. NLO) corrrections due to the Wilson coefficient~$C_{7}^{\mathrm{eff}}$
and in the $b\to s\gamma $ vertex, respectively, and $A_{\mathrm{sp}%
}^{(1)K_{1}}$ is the $\mathcal{O}(\alpha _{s})$ hard-spectator corrections
to the $B\to K_{1}\gamma $ amplitude computed in this paper. Their explicit
expressions are as follows: 
\begin{eqnarray}
A_{C_{7}}^{(1)}(\mu ) &=&\frac{\alpha _{s}(\mu )}{4\pi }\,C_{7}^{(1)\mathrm{%
eff}}(\mu ),  \label{eq:A1tb-C7} \\
A_{\mathrm{ver}}^{(1)}(\mu ) &=&\frac{\alpha _{s}(\mu )}{4\pi }\left\{ \frac{%
32}{81}\left[ 13C_{2}^{(0)}(\mu )+27C_{7}^{(0)\mathrm{eff}}(\mu
)-9\,C_{8}^{(0)\mathrm{eff}}(\mu )\right] \ln \frac{m_{b}}{\mu }\right.
\label{eq:A1tb-ver} \\
&-&\left. \frac{20}{3}\,C_{7}^{(0)\mathrm{eff}}(\mu )+\frac{4}{27}\left(
33-2\pi ^{2}+6\pi i\right) C_{8}^{(0)\mathrm{eff}}(\mu
)+r_{2}(z)\,C_{2}^{(0)}(\mu )\right\} ,\qquad  \nonumber \\
A_{\mathrm{sp}}^{(1)1^{1}P_{1}}(\mu _{\mathrm{sp}}) &=&\frac{\alpha _{s}(\mu
_{\mathrm{sp}})}{4\pi }\,\frac{2\Delta F_{\perp }^{(1^{1}P_{1})}(\mu _{%
\mathrm{sp}})}{9\xi _{\perp }^{(K_{1})}}\left\{ 3C_{7}^{(0)\mathrm{eff}}(\mu
_{\mathrm{sp}})\right.  \label{eq:A1tb-sp} \\
&+&\left. C_{8}^{(0)\mathrm{eff}}(\mu _{\mathrm{sp}})\left[ 1-\frac{%
6a_{\perp 1}^{(1^{1}P_{1})}(\mu _{\mathrm{sp}})}{\left\langle \bar{u}%
^{-1}\right\rangle _{\perp }^{(1^{1}P_{1})}(\mu _{\mathrm{sp}})}\right]
+C_{2}^{(0)}(\mu _{\mathrm{sp}})\left[ 1-\frac{h^{(1^{1}P_{1})}(z,\mu _{%
\mathrm{sp}})}{\left\langle \bar{u}^{-1}\right\rangle _{\perp
}^{(1^{1}P_{1})}(\mu _{\mathrm{sp}})}\right] \right\} .  \nonumber
\end{eqnarray}
Actually $C_{7}^{(1)\mathrm{eff}}(\mu )$ and $A_{\mathrm{ver}}^{(1)}(\mu )$
are process independent and encodes the QCD\ effects only, where as $A_{%
\mathrm{sp}}^{(1)}(\mu _{\mathrm{sp}})$ contains the key information about
the out going mesons. The factor $\frac{6a_{\perp 1}^{(1^{1}P_{1})}(\mu _{%
\mathrm{sp}})}{\left\langle \bar{u}^{-1}\right\rangle _{\perp
}^{(1^{1}P_{1})}(\mu _{\mathrm{sp}})}$ appear in the Eq. (\ref{eq:A1tb-C7})
is arising due to the Gegenbauer moments.

\begin{center}
$\stackunder{\text{Table1: Input quantities and their values used in the
theoretical analysis}}{
\begin{tabular}{|l|l|l|l|}
\hline
$\text{Parameters}$ & $\text{Values}$ & $\text{Parameters}$ & $\text{Values}$
\\ \hline
$M_{W}$ & $80.423\text{ GeV}$ & $M_{Z}$ & $91.1876\text{ GeV}$ \\ \hline
$M_{B}$ & $5.279\text{ GeV}$ & $m_{b_{1}}$ & $1.229$ \\ \hline
$G_{F}$ & $1.166\times 10^{-5}\text{ GeV}$ & $m_{h_{1}}$ & $1.170$ \\ \hline
$\alpha _{s}\left( M_{Z}\right) $ & $0.1172$ & $\alpha $ & $1/137.036$ \\ 
\hline
$m_{t\text{,pole}}$ & $178\text{ GeV}$ & $\Lambda _{h}$ & $0.5\text{ GeV}$
\\ \hline
$\left| V_{tb}V_{td^{*}}\right| $ & $5\times 10^{-3}$ & $m_{b\text{,pole}}$
& $4.27\text{ GeV}$ \\ \hline
$f_{B}$ & $200\text{ MeV}$ & $\sqrt{z}=m_{c}/m_{B_{d}^{0}}$ & $0.29$ \\ 
\hline
$a_{\perp 1}^{\left( b_{1}\right) }\left( 1\text{GeV}\right) $ & $0$ & $%
a_{\perp 2}^{\left( b_{1}\right) }\left( 1\text{GeV}\right) $ & $0.1$ \\ 
\hline
$a_{\perp 1}^{\left( h_{1}\right) }\left( 1\text{GeV}\right) $ & $0$ & $%
a_{\perp 2}^{\left( h_{1}\right) }\left( 1\text{GeV}\right) $ & $0.35$ \\ 
\hline
$f_{\perp }^{\left( b_{1}\right) }$ & $180\text{ MeV}$ & $f_{\perp }^{\left(
h_{1}\right) }$ & $200\text{ MeV}$ \\ \hline
$\lambda _{B\text{,}+}^{-1}$ & $\left( 2.15\pm 0.50\right) \text{ GeV}^{-1}$
& $\sigma _{B\text{,}+}\left( 1\text{ GeV}\right) $ & $1.4\pm 0.4$ \\ \hline
\end{tabular}
}$
\end{center}

\subparagraph{4.1. Branching ratios.}

\label{ssec:Branching-Ratios} We now proceed to calculate numerically the
branching ratios for the $B_{d}^{0}\to b_{1}\gamma $ and $B_{d}^{0}\to
h_{1}\gamma $ decays. The theoretical ratios involving the decay widths on
the r.h.s. of these equations can be written in the form: 
\begin{eqnarray}
R_{\mathrm{th}}(b_{1}\gamma /K_{1}\gamma ) &=&\frac{\mathcal{B}_{\mathrm{th}%
}(B_{d}^{0}\to b_{1}\gamma )}{\mathcal{B}_{\mathrm{th}}(B_{d}^{0}\to
K_{1}\gamma )}=\frac{1}{2}\left| \frac{V_{td}}{V_{ts}}\right| ^{2}\frac{%
(M_{B}^{2}-m_{b_{1}}^{2})^{3}}{(M_{B}^{2}-m_{K_{1}}^{2})^{3}}\,\zeta
^{2}\,\left[ 1+\Delta R(b_{1}/K_{1})\right] ,\qquad  \label{eq:Rth-rho/Ks} \\
R_{\mathrm{th}}(h_{1}\gamma /K_{1}\gamma ) &=&\frac{\overline{\mathcal{B}}_{%
\mathrm{th}}(B_{d}^{0}\to h_{1}\gamma )}{\overline{\mathcal{B}}_{\mathrm{th}%
}(B_{d}^{0}\to K_{1}\gamma )}=\frac{1}{2}\left| \frac{V_{td}}{V_{ts}}\right|
^{2}\frac{(M_{B}^{2}-m_{h_{1}}^{2})^{3}}{(M_{B}^{2}-m_{K_{1}}^{2})^{3}}%
\,\zeta ^{2}\,\left[ 1+\Delta R(h_{1}/K_{1})\right] ,\qquad
\label{eq:Rth-omega/Ks}
\end{eqnarray}
where~$m_{b_{1}}$ and~$m_{h_{1}}$ are the masses of the $b_{1}$- and $h_{1}$%
-mesons, $\zeta $~is the ratio of the transition form factors, which we have
assumed to be the same for the $b_{1}^{0}$- and $h_{1}$-mesons. To get the
theoretical branching ratios for the decays $B_{d}^{0}\to b_{1}\gamma $ and $%
B_{d}^{0}\to h_{1}\gamma $, the ratios~(\ref{eq:Rth-rho/Ks}) and~(\ref
{eq:Rth-omega/Ks}) should be multiplied with the corresponding experimental
branching ratio of the $B_{d}^{0}\to K_{1}\gamma $ decay.

It is well know that in vector meson case the theoretical uncertainty in the
evaluation of the $R_{\mathrm{th}}(\rho \gamma /K^{*}\gamma )$ and $R_{%
\mathrm{th}}(\omega \gamma /K^{*}\gamma )$ ratios is dominated by the
imprecise knowledge of $\zeta =\bar{T}_{1}^{\rho }(0)/\bar{T}_{1}^{K^{*}}(0)$
characterizing the $SU(3)$ breaking effects in the QCD transition form
factors. In the $SU(3)$-symmetry limit, $\bar{T}_{1}^{\rho }(0)=\bar{T}%
_{1}^{K^{*}}(0)$, yielding $\zeta =1$. We make use of the $SU\left( 3\right) 
$ symmetry to relate the form factor of $B\to b_{1}\gamma $ and $%
B_{d}^{0}\to h_{1}\gamma $ with that of $B_{d}^{0}\to K_{1}\gamma $ decay
which is the only unknown parameter involved in the calculation of branching
ratio for these decays. We use this symmetry because there is no
experimental limit on the branching ratio of these decays. It is reasonable
to use $\xi _{\perp }^{1^{1}P_{1}}(0)=\xi _{\perp }^{K_{1}}(0)$ because $%
SU\left( 3\right) $ symmetry is good for the form factors irrespective of
the fact that it is not exact for the masses.~Thus in present analysics we
use $\xi _{\perp }^{(b_{1}\text{, }h_{1})}=0.32$ together with the values of
the other input parameters entering in the calculation of the $B_{d}^{0}\to
(b_{1},h_{1})\,\gamma $ decay amplitudes and these are given in Table 1.

The individual branching ratios $\mathcal{B}_{\mathrm{th}}(B_{d}^{0}\to
b_{1}\gamma )$ and $\mathcal{B}_{\mathrm{th}}(B_{d}^{0}\to h_{1}\gamma )$
and their ratios $R_{\mathrm{th}}(b_{1}\gamma /K_{1}\gamma )$ and $R_{%
\mathrm{th}}(h_{1}\gamma /K_{1}\gamma )$ with respect to the corresponding $%
B\to K_{1}\gamma $ branching ratios are calculated and the corresponding
values are: 
\begin{eqnarray}
\mathcal{B}_{\mathrm{th}}[B_{d}^{0} &\to &b_{1}\gamma ]=0.71\times 10^{-6}
\label{eq:Brth-Rho-average} \\
\mathcal{B}_{\mathrm{th}}[B_{d}^{0} &\to &h_{1}\gamma ]=0.74\times 10^{-6}
\label{eq:Brth-Rho-average-1} \\
\mathcal{R}_{\mathrm{th}}[b_{1}\gamma /K_{1}\gamma ] &=&0.0166
\label{eq:Rth-Rho-average} \\
\mathcal{R}_{\mathrm{th}}[h_{1}\gamma /K_{1}\gamma ] &=&0.0167
\label{eq:Rth-Rho-average1}
\end{eqnarray}
To calculated these values we have used the experimental value of the
branching ratio of $B\to K_{1}\gamma $. One can eaisly see that there is
very small difference between $B_{d}^{0}\to b_{1}\gamma $ and $B_{d}^{0}\to
h_{1}\gamma $ branching fractions, and this is due to the slight change in
the hadronic parameters of these decays.

The $SU(3)$-breaking effects in $\rho $ and $K^{*}$ form factors 
have been evaluated within several approaches, including the LCSR and
Lattice QCD. In the earlier calculations of the ratios~for $B\to \rho \gamma 
$ and $B\to K^{*}\gamma $\cite{Ali:2001ez,Ali:2002kw}, the following ranges
were used: $\zeta =0.76\pm 0.06$~\cite{Ali:2001ez} and $\zeta =0.76\pm 0.10$~%
\cite{Ali:2002kw}, based on the LCSR approach~\cite
{Ali:vd,Ali:1995uy,Ball:1998kk,Narison:1994kr,Melikhov:2000yu} which
indicate substantial $SU(3)$ breaking in the $B\to K^{*}$ form factors.
There also exists an improved Lattice estimate of this quantity, $\zeta
=0.9\pm 0.1$\cite{Becirevic:2003}. To incorporate the $SU\left( 3\right) $
symmetry for these axial meson decays we have plotted the branching ratios
of $B_{d}^{0}\to (b_{1},h_{1})\,\gamma $ decay with the LEET\ form factor
which is presented in Fig.1. The solid and dashed line show the dependence
of the branching ratio of $B_{d}^{0}\to b_{1}\,\gamma $ and $B_{d}^{0}\to
h_{1}\,\gamma $ on the LEET form factor $\xi _{\perp }^{1^{1}P_{1}}(0)$
respectively. The graph shows that in the range $0.76\leq \zeta \leq 0.9$
the value of the branching ratio $\left( \text{in the units of }%
10^{-6}\right) $ is $0.4\leq \left( B_{d}^{0}\to 1^{1}P_{1}\gamma \right)
\leq 0.7$.

\paragraph{4.2. CP-violating asymmetries.}

\label{ssec:CP-asymmetry} The direct CP-violating asymmetries in the decay
rates for $B_{d}^{0}\to (b_{1},h_{1})\,\gamma $ decays are defined as
follows: 
\begin{eqnarray}
\mathcal{A}_{\mathrm{CP}}^{\mathrm{dir}}(b_{1}\gamma ) &\equiv &\frac{%
\mathcal{B}(\bar{B}_{d}^{0}\to b_{1}\gamma )-\mathcal{B}(B_{d}^{0}\to
b_{1}\gamma )}{\mathcal{B}(\bar{B}_{d}^{0}\to b_{1}\gamma )+\mathcal{B}%
(B_{d}^{0}\to b_{1}\gamma )},  \label{eq:CPasym-def} \\
\mathcal{A}_{\mathrm{CP}}^{\mathrm{dir}}(\omega \gamma ) &\equiv &\frac{%
\mathcal{B}(\bar{B}_{d}^{0}\to h_{1}\gamma )-\mathcal{B}(B_{d}^{0}\to
h_{1}\gamma )}{\mathcal{B}(\bar{B}_{d}^{0}\to h_{1}\gamma )+\mathcal{B}%
(B_{d}^{0}\to h_{1}\gamma )}.  \nonumber
\end{eqnarray}
Before we go for the numerical values of CP-asymmetry, let us discuss the
difference in the hadronic parameters involving the $b_{1}$ and $h_{1}$
mesons. As these are the axial vector states of $\rho ^{0}$ and $\omega $
mesons so these are also the maximally mixed superpositions of the~$\bar{u}u$
and~$\bar{d}d$ quark states: $|b_{1}\rangle =(|\bar{d}d\rangle -|\bar{u}%
u\rangle )/\sqrt{2}$ and $|h_{1}\rangle =(|\bar{d}d\rangle +|\bar{u}u\rangle
)/\sqrt{2}$. Neglecting the $W$-exchange contributions in the decays, the
radiative decay widths are determined by the penguin amplitudes which
involve only the~$|\bar{d}d\rangle $ components of these mesons, leading to
identical branching ratios (modulo a tiny phase space difference). The $W$%
-exchange diagrams from the~$\mathcal{O}_{1}^{(u)}$ and~$\mathcal{O}%
_{2}^{(u)}$ operators (in our approach, we are systematically neglecting the
contributions from the penguin operators $\mathcal{O}_{3},...,\mathcal{O}%
_{6} $) yield contributions equal in magnitude but opposite in signs[for
detailed calculation please see \cite{Alinew,Grinstein:2000pc}]. If we use
the notations and expressions given in Ref.\cite{Grinstein:2000pc}, the LCSR
results are: $\epsilon _{A}^{(b_{1})}=$ $-\epsilon _{A}^{(h_{1})}=0.07$~.
The smallness of these numbers reflects both the colour-suppressed nature of
the $W$-exchange amplitudes in $B_{d}^{0}\to (b_{1}$,$h_{1})\,\gamma $
decays, and the observation that the leading contributions in the weak
annihilation and $W$-exchange amplitudes arise from the radiation off the $d$%
-quark in the $B_{d}^{0}$-meson, which is suppressed due to the electric
charge.

The explicit expressions of these asymmetries for the charged axial vector
meson in terms of the individual contributions in the decay amplitude can be
found in Ref.\cite{jamilnew}, which for $\mathcal{A}_{\mathrm{CP}}^{\mathrm{%
dir}}(b_{1}\gamma )$ and $\mathcal{A}_{\mathrm{CP}}^{\mathrm{dir}%
}(h_{1}\gamma )$ may be obtained by obvious replacements. The calculated
values of the CP-asymmetry for the above mentined decays are summarized in
Table 2. The CP-asymmetry recieves contributions from the hard spectator
corrections which tend to decrease its value estimated from the vertex
corrections alone. The dependence of the direct CP-asymmetry on the CKM
unitarity-triangle angle $\alpha $ is presented in the Fig.2. It should be
noted that the predicted direct CP-asymmetries are rather sizable (of
order~10\%) and is negative like $\rho $ and $\omega $ meson case. It is
quite unfortunate that the predicited value of CP asymmetry is sensitive to
both the choice of the scale as well as the quark mass ratio $%
z=m_{c}^{2}/m_{b}^{2}$ used in the calculation.

\begin{center}
\begin{tabular}{|l|l|l|}
\hline
& $B_{d}^{0}\to b_{1}\gamma $ & $B_{d}^{0}\to h_{1}\gamma $ \\ \hline
$\mathcal{R}_{\mathrm{th}}$ & $0.0166$ & $0.0167$ \\ \hline
$\mathcal{B}_{\mathrm{th}}$ & $0.71\times 10^{-6}$ & $0.74\times 10^{-6}$ \\ 
\hline
$\mathcal{A}_{CP}^{\text{dir}}$ & $-10.7\%$ & $-9.8\%$ \\ \hline
\end{tabular}
\end{center}

\paragraph{5. Summary}

\label{sec:summary} We have calculated the branching ratios for $%
b\rightarrow 1^{1}P_{1}\gamma $ decays at NLO of $\alpha _{s}$. These $%
1^{1}P_{1}$ are $b_{1}$ and $h_{1}$ mesons which are the corresponding
radially excited axial vector mesons of $\rho $ and $\omega $ respectively.
Using the $SU\left( 3\right) $ symmetry for the form factor, the branching
ratio for $B_{d}^{0}\rightarrow \left( b_{1},h_{1}\right) \gamma $ is
expressed in terms of the branching ratio of the $B_{d}^{0}\rightarrow
K_{1}\gamma $ and it is found to be $\mathcal{B}\left( B_{d}^{0}\rightarrow
b_{1}\gamma \right) =0.71\times 10^{-6}$ and $\mathcal{B}\left(
B_{d}^{0}\rightarrow h_{1}\gamma \right) =0.74\times 10^{-6}$. Then we have
plotted the branching ratio with the LEET form factor which is the only
unknown parameter involved in the calculation. It is shown that the
corresponding to the range of $SU\left( 3\right) $ symmetry breaking
parameter$\zeta $ , $0.76\leq \zeta \leq 0.9$ the value of the branching
ratio $\left( 10^{6}\right) $ is $0.4\leq \left( B_{d}^{0}\to
1^{1}P_{1}\gamma \right) \leq 0.7$. Therefore in future when we have the
experimental data on these decays we will be able to extract the value of
form factor. Further we have also calculated direct CP asymmetry for these
decays and find, in confirmity with the observations made in the literature,
that the hard spectator contributions significantely reduces the asymmetry
arising from the vertex corrections alone. The value of CP-asymmetry is 10\%
and is negative like $\rho $ and $\omega $ in the Standard Model. Thus the
measurement of CP-asymmetry will either overconstrain the angle~$\alpha $ of
the unitarity triangle, or they may lead to the discovery of physics beyond
the~SM in the radiative $b\to d\gamma $ decays.

\paragraph{Acknowledgements.}

One of the authors (J) would like to thank Prof. Fayyazuddin for valuable
discussion.This work was supported by a grant from Higher Education
Commission of Pakistan.

\textbf{Figure Captions}

Figure1: Branching ratio for $B\rightarrow 1^{1}P_{1}\gamma $ decay vs LEET
form factor; Solid line shows the value for $b_{1}$ meson and the dashed
line is for $h_{1}$ meson.

Figure2: CP-asymmetry ($-\mathcal{A}_{CP}\%$) vs the Unitarity triangle
phase $\alpha $; Solid line is for $h_{1}$ meson and dashed line is for $%
b_{1}$ meson.

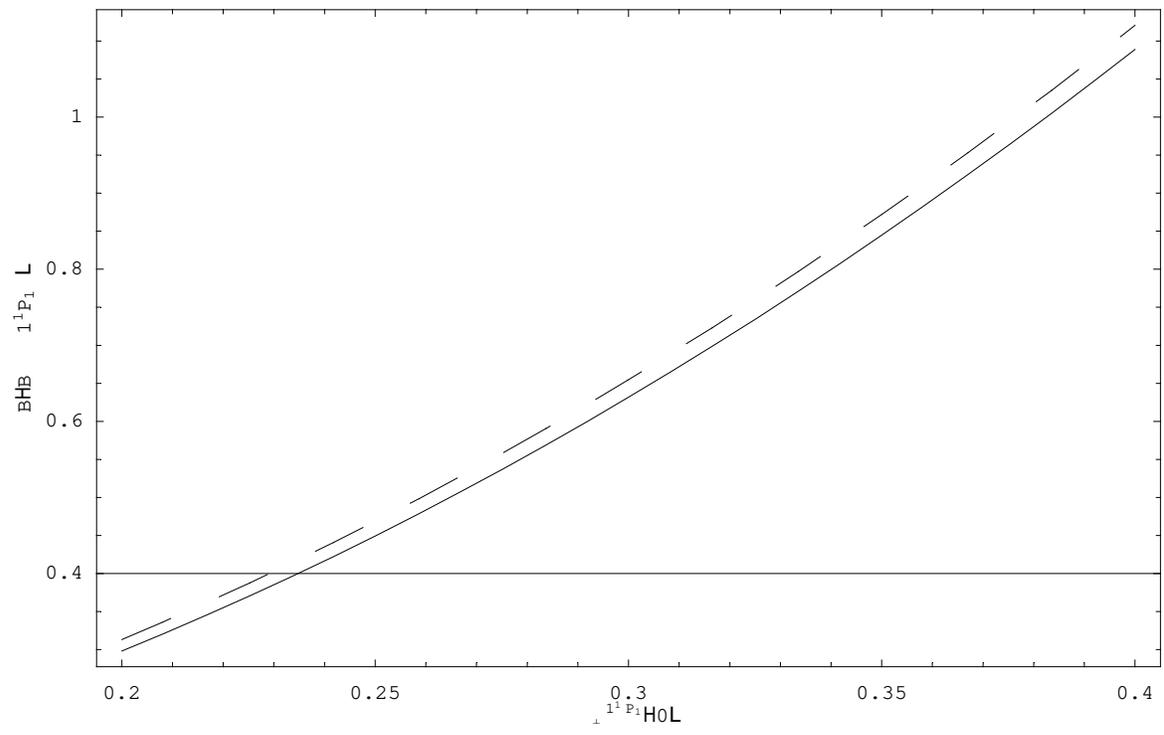
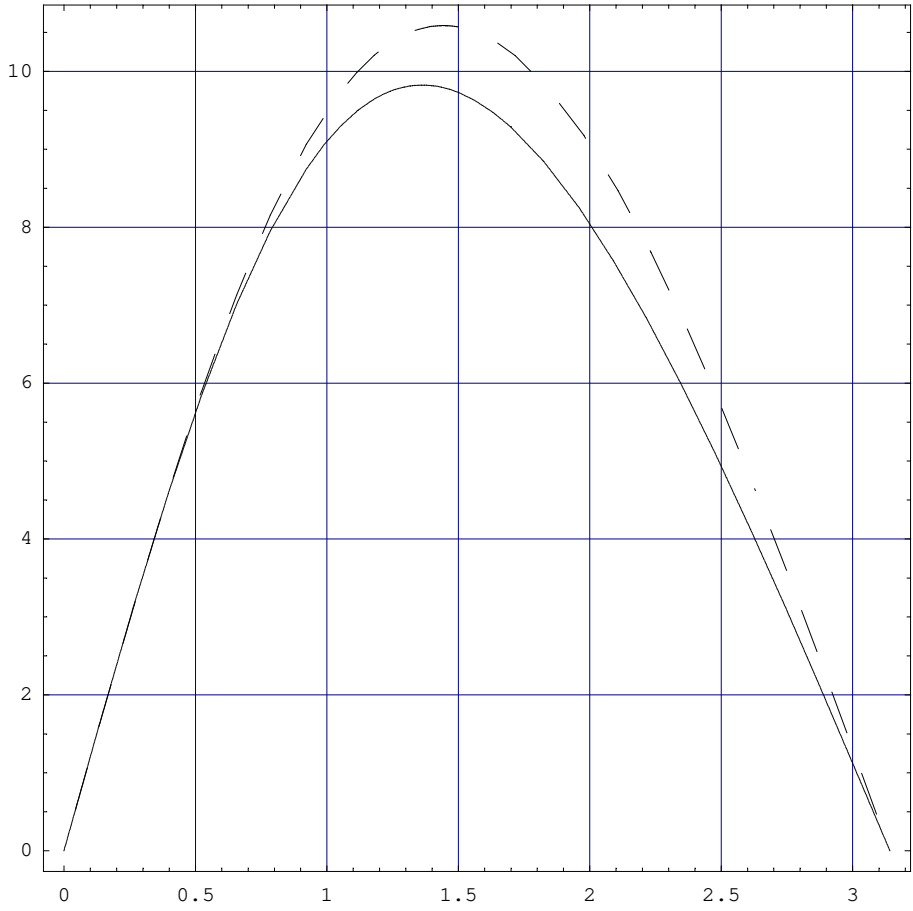
\end{document}